# Lost Vibration Test Data Recovery Using Convolutional Neural Network: A Case Study

P. Moeinifard, M. S. Rajabi, M. Bitaraf *

School of Civil Engineering, College of Engineering, University of Tehran, Tehran, Iran.

**ABSTRACT:** Data loss in Structural Health Monitoring (SHM) networks has recently become one of the main challenges for engineers. Therefore, a data recovery method for SHM, generally an expensive procedure, is essential. Lately, some techniques offered to recover this valuable raw data using Neural Network (NN) algorithms. Among them, the convolutional neural network (CNN) based on convolution, a mathematical operation, can be applied to non-image datasets such as signals to extract important features without human supervision. However, the effect of different parameters has not been studied and optimized for SHM applications. Therefore, this paper aims to propose different architectures and investigate the effects of different hyperparameters for one of the newest proposed methods, which is based on a CNN algorithm for the Alamosa Canyon Bridge as a real structure. For this purpose, three different CNN models were considered to predict one and two malfunctioned sensors by finding the correlation between other sensors, respectively. Then the CNN algorithm was trained by experimental data, and the results showed that the method had a reliable performance in predicting Alamosa Canyon Bridge's missed data. The accuracy of the model was increased by adding a convolutional layer. Also, a standard neural network with two hidden layers was trained with the same inputs and outputs as the CNN models. Based on the results, the CNN model had higher accuracy, lower computational cost, and was faster than the standard neural network.

## 1- Introduction

Available infrastructures are subjected to considerable operational and environmental loads during their life cycle. These loads and hazardous events like earthquakes and extreme hurricanes can impact structures unfavorably and speed up structural damage. Hence, identifying the structural condition timely is essential to ensure safety [1]. Traditionally, visual inspection was crucial in evaluating the structural condition. Nevertheless, visual inspection is w=ork-intensive and time-wasting; consequently, it cannot accurately track condition variations in real-time. Structural health monitoring (SHM) approaches emerged to address this issue and acquired increasing usage in past decades. SHM is a useful method that presents tools for evaluating and monitoring structural health [2, 3]. These tools can widely be used for ensuring integrity and safety, detecting damage, and estimating performance deterioration of infrastructures [3-8].

During the last 40 years, diverse SHM methods have been increasingly exploited to monitor bridges' structural conditions around the world [9-12]. These methods have collected immense data after long-term data gathering time by numerous sensors with high sampling frequencies [13]. In current SHM applications, the utilization of these sensors faces lots of limitations, including sensor measurement disabilities or sensor failure, which can be momentary or permanent [13-15]. Also, there are cases in which data loss may occur during data transmission between sensors and receivers in wireless SHMS [16]. These limitations controlled the efficient use of dense sensor arrays on civil infrastructure and resulted in insufficient datasets, which caused a significant problem in starting the analysis phase due to created missing data [17]. Furthermore, considering the limitations of computational ability and data analysis methods, the knowledge of a large amount of SHM data is not well interpreted. Big data (BD) and artificial intelligence (AI) techniques are discussed as advantageous practices to address the issues mentioned earlier [13].

Several methods have been conducted to recover the missing data resulting from sensors or transition failures during recent years. These recovering methods are mainly based on estimating those data using other data collected from different sensors [18]. In a study, recovering of missed data was performed by a compressive sampling technique using fewer measured data compared to the number of measurement data made obtainable by SHM [19]. This compressive technique is also employed to recover dynamic structural responses in a wireless network for long-term structural monitoring of a bridge [20]. Data recovery methods have also been applied to recover missing data. In a study, restoring the missing data was achieved by interpolating the characteristics between the data measured by stress sensors installed on the building's structural members [15]. Lu *et al.* employed the partial least square method to propose an approach to recovering strain monitoring data in buildings. In this method, the correlation between the strain sensors installed at different building locations has been proved by a proposed model [21]. Also, an inter-sensor modeling method was offered in a study, to recover strain data collected from sensors of an actual bridge using nonparametric copulas. During the study, the basis of predicting data was the dependence of sensors located at different locations on the bridge [22].

*Corresponding author's email: maryam.bitaraf@ut.ac.ir

Deep neural networks (DNN) are other tools to recover and predict missing data in SHM networks [23, 24]. Liu *et al.* utilized a deep learning-based data recovery method for missing structural temperature data in an SHMS mounted on a river bridge. Based on their results, they found that the long short-term memory (LSTM) network-based recovery method performs better in terms of accuracy than the support vector machine (SVM), and wavelet neural network (WNN) in predicting time series data such as structural temperature [18]. In a recent study, to detect missing dynamic responses, a DNN was proposed and was trained using visualized time series structural responses and missing data from dynamic accelerations measured from long-term SHM [25]. In another study, a novel neural network-based method has been employed to predict high-rise buildings' future structural responses using the previous and current steps of the measured time history acceleration data [26]. In another study, a NN was proposed to estimate responses of buildings proposed. In this study, the input included building structural parameters and the wind load, and output was set as the maximum inter-story drift ratio [27]. Predicting large buildings' structural responses was also studied in another study using a recurrent neural network (RNN). Perez-Ramirez *et al.* proposed an RNN model in which the acceleration time history responses of particular floors in a structure were assumed as the input layer, and that of another story was set as the output layer of NN. This model's prediction was examined and verified by structural responses under various lateral loads [28].

In recent years, CNN is broadly utilized in image recognition, image classifications, etc. To move toward automated monitoring, CNNs are progressively employed by detecting structural flaws in images. CNNs can address the most traditional methods' weaknesses, which require hand-crafted features. One of the main features of CNNs is being qualified for negotiating with an ample amount of data. Furthermore, recent applications of CNNs show that these models are powerful in solving overfitting problems, which are counted as the most critical problem of conventional NNs. In the SHM fields, the CNNs also show excellent performances in dealing with pixel-based image data. These applications mostly can be seen to address image-based damage detection, corrosion detection, and visual-based damage identification [29-31]. A method for crack identification using large-scale pixel-based images has been introduced by Xu *et al.* These images were classified into sub-images, and the proposed CNNs were trained with these sub-images. Using new images taken from girders' surfaces in bridges, the CNN model automatically identified cracks [29]. Khodabandehlou *et al.* employed the image classification ability of CNNs to introduce a method for evaluating structural conditions [32]. In another research, the proposed CNN model was used to detect data anomaly that usually occurs in signal processing of measured data gathered from field tests. In this approach, time and frequency information of dynamic structural responses were set as the trained CNN model's input for visualization of image data, and these data were classified as normal and anomalies [33]. Fan *et al.* proposed a data recovery method for vibration responses of bridges based on deep learning. In this technique, measured responses from structures were used as input and output of CNNs [34]. Oh *et al.* utilized a CNN-based data recovery method in the case of a sensor fault or data loss in a sensor network. This method recovers the missing strain data using the remaining sensors' responses by training the CNN measured before the failure of the mentioned sensor [17].

Although some previous research implemented ML algorithms to recover the missed data laboratory specimens, the reliability of this method has not been studied for the real and in-site structures such as bridges or existing raw data of SHM. Besides, the effects of different parameters have not been studied [35]. Therefore, in this paper, a recovery method using CNN was utilized to predict the missing data of accelerometers installed on Alamosa Canyon Bridge in the case of structural health monitoring [36]. Then modal features such as natural frequencies and mode shapes were then obtained using the predicted data and compared with the original data. Moreover, the effect of different parameters such as activation function, number of layers, etc., was investigated, and a comparison was made between CNN and NN. Finally, a standard neural network (NN) was implemented to compare it to CNN which has not been studied so far.

This paper is organized as follows: In section 2, brief information was provided about the Alamosa Canyon Bridge studied in this paper. Then, in section 3, to implement the data recovery method, different architectures for CNN and NN were proposed with their formulas. In section 4, the result of the models was shown as the recovered data of faulted sensors and modal parameters. Besides, an error analysis was done. Furthermore, in section 5, a parametric study was done to obtain the best activation function. Finally, in the last section, the conclusion of this study was provided.

## 2- General overview of the bridge

Alamosa Canyon Bridge (Fig. 1) was a steel bridge with a concrete deck that was located from north to south in New Mexico, USA. The bridge was reconstructed in the 1960s based on seven independent spans. It used six steel W30 × 116 standard wide-flange beams which had roller connections at their ends to transfer loads of each span to the piers. Also, the cross braces of this bridge were channel sections (C12 × 25). Fig. 2 schematically shows the elevation view of the Alamosa Canyon Bridge and its dimension.

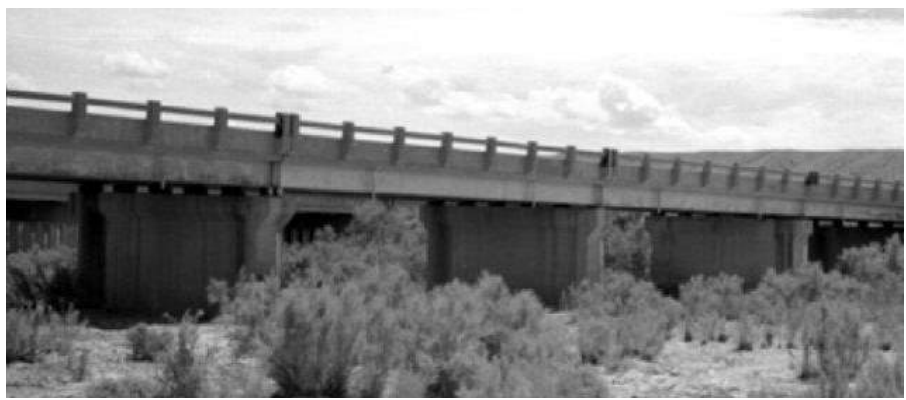

**Fig. 1. Alamosa Canyon Bridge [37].**

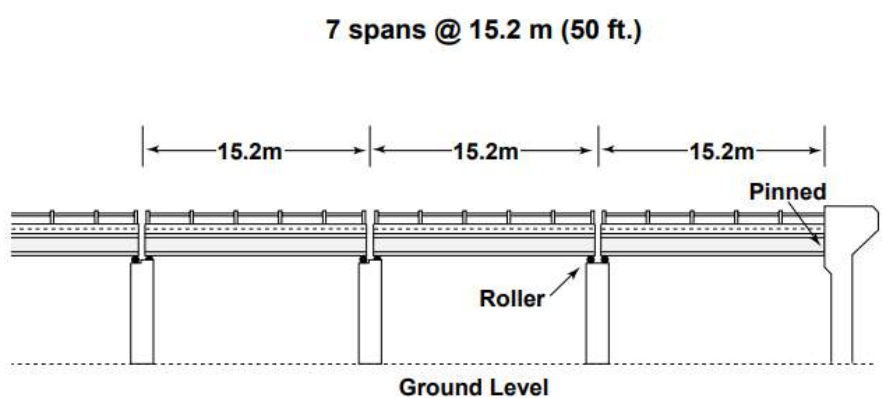

**Fig. 2. Side view of the Alamosa Canyon Bridge [36].**

**Table 1. Summarized field tests of the Alamosa Canyon Bridge.**

| Structural Health Monitoring test | Vibration test | Type of sensors | Number of data acquisition channels | Number of tests |
|---|---|---|---|---|
| Forced vibration test | 24hr Impact Test | Accelerometer and Transducer | 32 | 11 |
| | Random shaker test | Accelerometer and Transducer | 32 | 10 |
| Ambient Vibration Test | Ambient Vibration Test | Accelerometer and Transducer | 32 | 11 |

## 2- 1- Structural health monitoring system

In the 1990s, Farrar *et al.* [36] conducted several field tests on Alamosa Canyon Bridge. The main purpose of those tests, which were funded by Los Alamos National Laboratory's Laboratory Direct Research and Development (LDRD) office, was to investigate damage detection methods in situ structures. Table 1 summarizes the field tests which were used in the previous research to validate damage detection meth-ods. Several global damage detection methods were studied. Also, they investigated how modal parameters would change under different environmental and operational conditions.

On each test that has been conducted by Farrar [36] on Alamosa Canyon Bridge, different numbers of accelerometers have been installed. Fig. 3 shows the location of installed sensors on the Alamosa Canyon Bridge for an ambient test, which is studied in this paper.

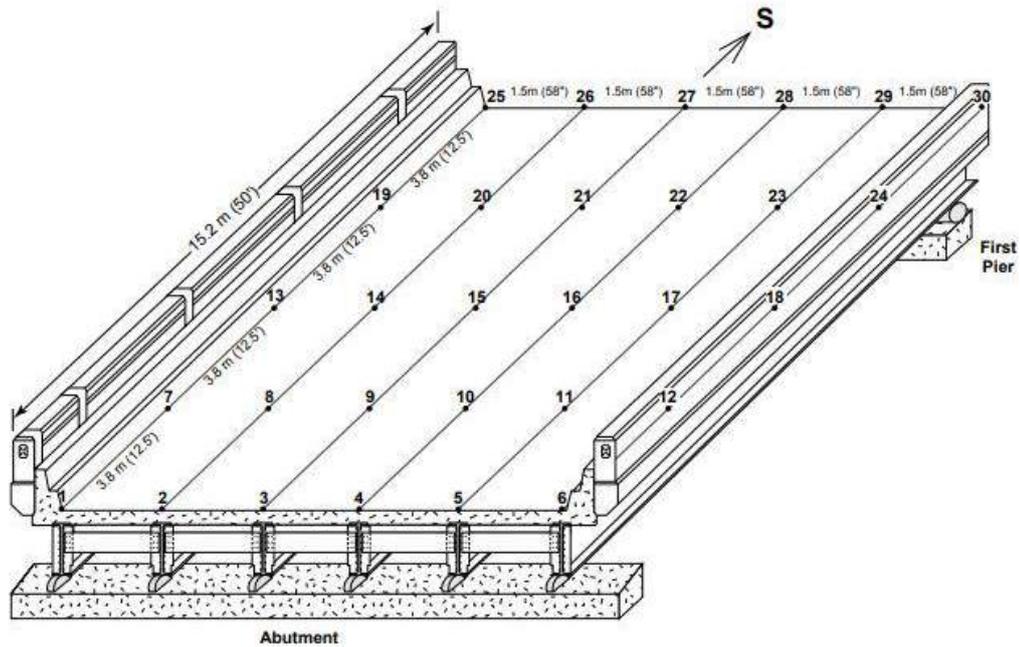

**Fig. 3. Schematic location of accelerometers in Farrar's experiment [36].**

### 3- Materials and Methods

The convolutional neural network, which is usually called CNN or ConvNet, is one of the machine learning algorithms. CNN is mainly used for computer vision purposes such as image classification, face recognition, and so on. However, CNN can be applied to non-image datasets, especially waveform datasets such as audio and other types of signals. A CNN is based on a mathematical operation or sliding dot product. Each CNN architecture is formed by a series of layers that pass their results to the next layer.

#### 3- 1- Convolutional Neural Network

At first, in this study, all the recorded data of a test, an ambient test in this study, was assumed to be a matrix. Fig. 4 shows the procedure of arrangement and making the matrix. Each column of this matrix belongs to a sensor, and each row represents measured acceleration on each time step. The time steps of the studied data were 0.007812 seconds, and 2048 accelerations had been recorded. The total duration of each test was 16 seconds.

To decrease the training time of the proposed algorithm, which is one of the critical challenges in machine learning, it is essential to normalize and rescale the accused data before training the CNN model. Another reason for normalization was the different range of recorded data by sensors due to their various distance from the excitation source. Therefore, each sensor's data was normalized by finding the minimum and maximum of data and using the following equation:

$$Nd = \frac{Acc - Min}{Max - Min} \tag{1}$$

where *Acc* is each recorded acceleration in the dataset, *Max* stands for a maximum value of the recorded data, *Min* is the minimum value of the recorded acceleration, and *Nd* indicates the normalized values of raw data from experiments.

For the proposed method, to recover a specific sensor's data, first, it should be intentionally assumed as a malfunctioned or faulted sensor. When a sensor becomes faulted, it will not be able to measure the acceleration, and it can be assumed to record the zero as the magnitude of the acceleration during the malfunctioned period. Hence, its data was not considered and then replaced with zero in the matrix of raw data. Due to limitations on the available recorded data, to increase the number of training sets for the machine learning algorithm, data augmentation was implemented using overlapping the existing dataset, increasing the number of training sets. Hence, the dataset was divided into square windows to set as input. The windows' dimension was considered equal to the number of sensors. In the case of Fig. 1, the windows' dimension was 30*30. The total number of windows (*W*) is calculated as follows:

$$W = \frac{Nt - WL}{WS} \tag{2}$$

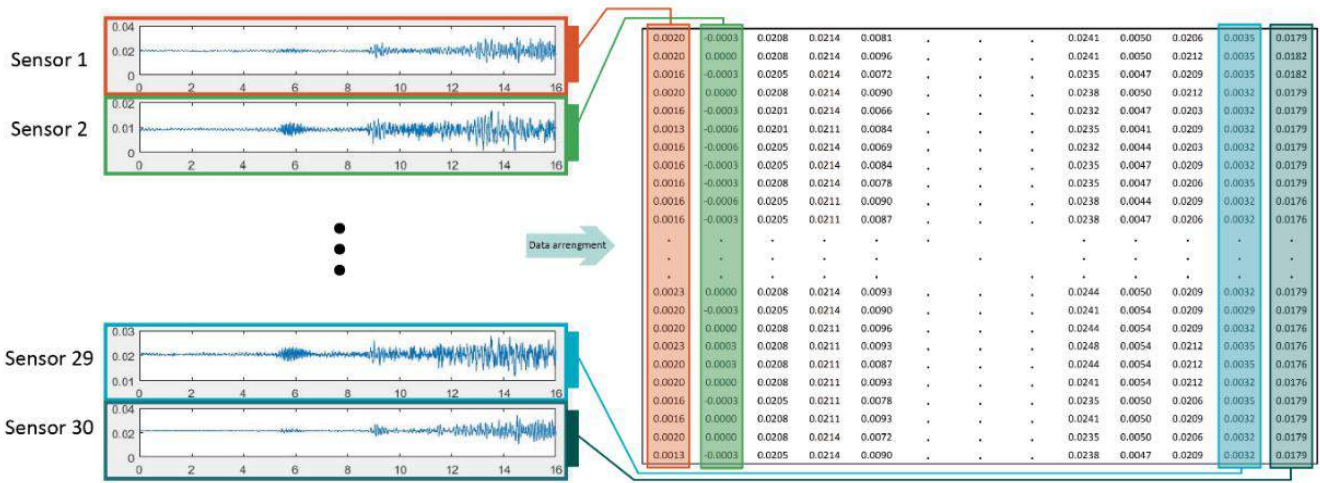

**Fig. 5. The proposed CNN architecture of model (a) for the recovery of Alamosa Canyon Bridge.**

*Nt*, *WL* and *WS* indicate the number of recorded accelerations, windows' length, and windows' stride, respectively. In this study, the *WL* was assumed to be equal to 30 (number of sensors). The outputs of the CNN are the data that have been omitted and replaced with zero. Decreasing the parameter *WS* provides more input data and will increase the accuracy of the prediction. Therefore, a sensitivity analysis was done, and the results showed that using values for *WS* smaller than 6 increased the computational costs; however, the error changes were not significant (less than 3%). Hence, the *WS* was assumed 6. Eighty percent of divided windows randomly were assumed as the training sets, and the others were used as validation sets. In this paper, three different CNN models have been considered, named (a), (b), and (c). Model (a) and (b) had 3 and 2 convolutional layers, respectively. They were trained to predict one faulted sensor's data. Model (c) had three convolutional layers; however, it was used and trained to recover two faulted sensors. Other hyperparameters, such as the number of kernels, strides, learning rate, and so on, were the same for these three models. Fig. 5 illustrates the proposed CNN architecture of model (a) for the data recovery method in this paper. The input datasets, which are arranged windows, pass through the first convolutional layer. This layer convolves inputs using 32 kernels, which are 8*8 matrix. These kernels were assumed as hyperparameters, and the algorithm tried to learn them. The convolutional layer's size depends on the input layer, kernel size, and stride sizes which are calculated as follows:

$$C = \frac{I - K}{Stride} + 1 \qquad (3)$$

Where the *C*, *I*, and *K* indicate the size of the convolutional layer, input layer, and kernel, respectively, and Stride indicates the number of data that the kernel shifts over the input layer each time.

The next layer is the pooling layer. This layer reduces the dimension of the convolutional layer's output by subsampling. In this paper, the max-pooling is used, which returns the maximum values of its input's rectangular regions. This layer is connected to the dropout layer, which is used as regularization to prevent overfitting. In the proposed model, these layers were repeated three times to increase the model's complexity and accuracy. Also, the number of filters doubles in each convolutional layer compared to the previous one. The last dropout layer flattened and connected to the dense layer (fully connected layer), which used the sigmoid function as its activation function. The details of the proposed CNN for the Alamosa Canyon Bridge are shown completely in Table 2.

The output of the dense layer is a 30*1 vector that returns the predicted data for a sensor which were intentionally assumed as a faulted sensor at the first step. The length of this vector depends on the windows' size, which was 30*30 Metrix in this study. After training the CNN with the training datasets, the validation data sets were used to evaluate the performance of the algorithm.

### 3- 2- Neural Network

To compare the advantage of the CNN algorithm for recovering data to other machine learning algorithms, a standard neural network was proposed and trained in this research. Fig. 6 shows the architecture of the NN. This NN consists of 2 hidden layers with the exponential linear unit (eLU) activation function, the same as CNN models. The training sets and validation sets were the same as CNN.

**Table 2. Details of CNN architecture for the first model (model (a)) to predict one sensor**

| Layer | Size | Operator | Size/Stride size/Number of kernels |
|---|---|---|---|
| Input layer | 30×30 | Kernel 1 | 17×17/1/32 |
| Convolutional layer 1 | 14×14×32 | Subsampling 1 | 2×2/1/- |
| Pooling layer 1 | 13×13×32 | Kernel 2 | 8×8/1/64 |
| Convolutional layer 2 | 6×6×64 | Subsampling 2 | 2×2/1/- |
| Pooling layer 2 | 5×5×64 | Kernel 3 | 4×4/1/128 |
| Convolutional layer 3 | 2×2×128 | Subsampling 3 | 2×2/1/- |
| Pooling layer 3 | 1×1×128 | | |
| FC layer | 128×1 | | |
| Output layer | 30×1 | | |

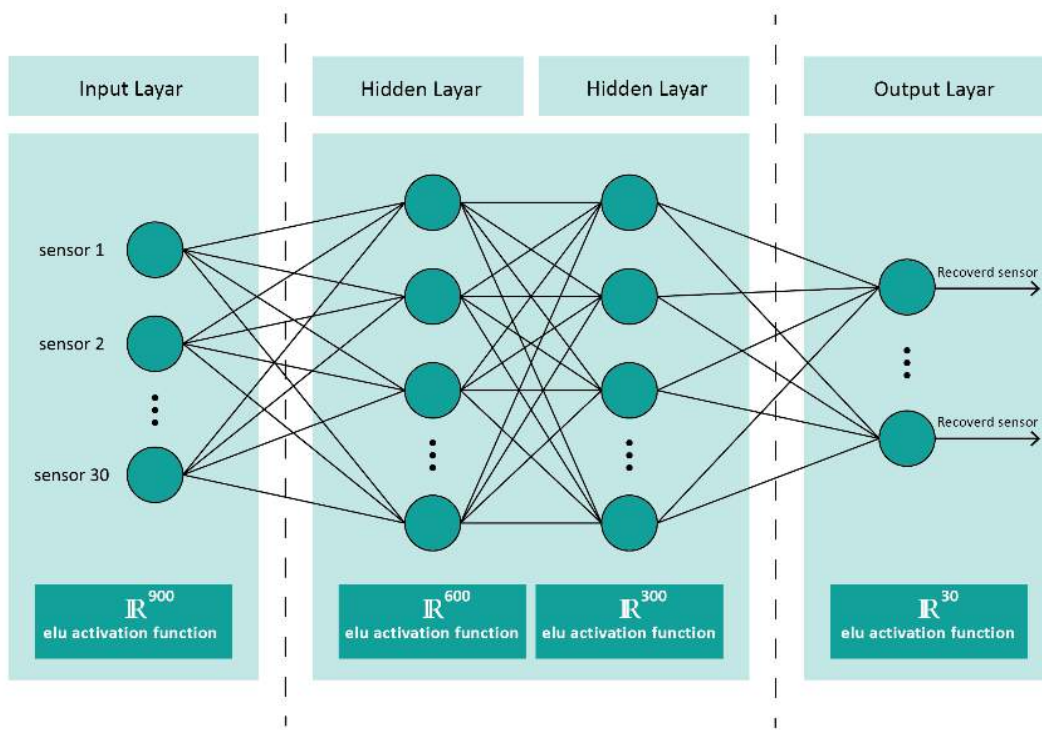

**Fig. 6. The proposed NN architecture for the recovery of Alamosa Canyon Bridge.**

All the accelerometers' responses at a window are passed as inputs, and the outputs are the responses of faulted sensors. The learning rate, the number of training, and other hyperparameters are assumed equal to CNN to have a fair conclusion.

## 4- Results and Discussion
### 4- 1- Regenerated Lost Signals
After the models were trained, four random periods (window), which were 0.234 seconds of the experimental test and contained 30 records of acceleration, were selected, and a damage scenario was assumed. For models (a) and (b), one sensor was assumed malfunctioned, and their records converted to zero. Then, the trained model predicted the real value of a malfunctioned sensor. Figs. 7 & 8 compare the normalized predicted data and the reference data for models (a) and (b), respectively. As seen, the predicted data are very close to reference data and accurately follow the entire data trend. As expected, model (a) (Fig. 7) showed better performance than model (b) (Fig. 8) due to having more convolutional layers and being more convoluted. But, model (b) (Fig.

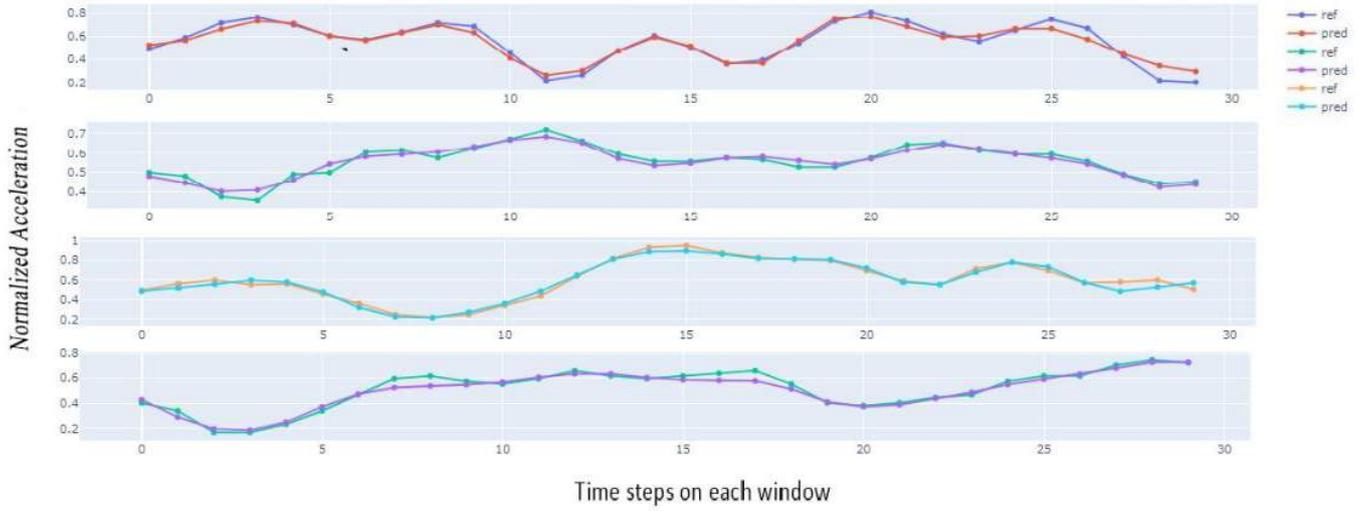

**Fig. 7. Acceleration prediction of the model (a) and the real values.**

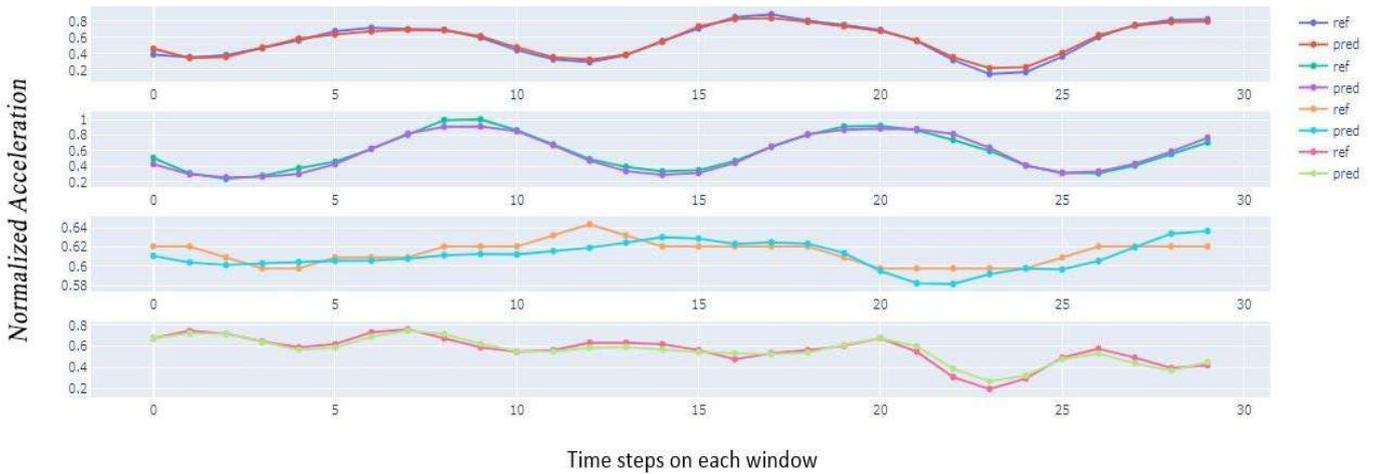

**Fig. 8. Acceleration prediction of the model (b) and the real values.**

8) did not show very accurate performance in the third period (window). The maximum error in the tested periods for models (a) and (b) were 2.9 and 3.12%, respectively.

For model (c), which was assumed to have two malfunctioned sensors, all the recorded data for these two sensors were assumed to zero in the input data. The output is a 60×1 array. Fig. 9 shows the results of model (c) in the abovementioned period. As shown, the first thirty data belong to the first sensor, and the rest belongs to the second malfunctioned sensor. The results show that this method exhibited desirable performance in prediction data of more than one sensor. The maximum difference between the reference and predicted data in the tested four periods was 4.67%.

Fig. 10 presents the results obtained from the conventional neural network. As shown, the proposed NN model could not predict the missed data very accurately compared to CNN models. However, the NN is still a reliable algorithm for use in the proposed method, and the maximum error in these four periods is less than 8% (7.4%).

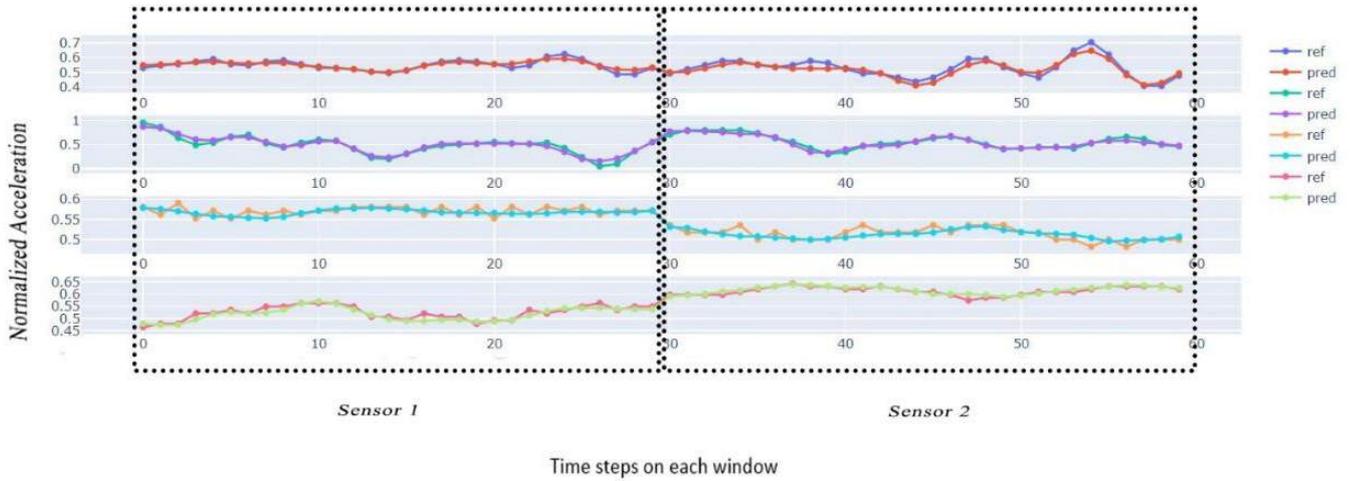

**Fig. 9. Acceleration prediction of the model (c) and the real values.**

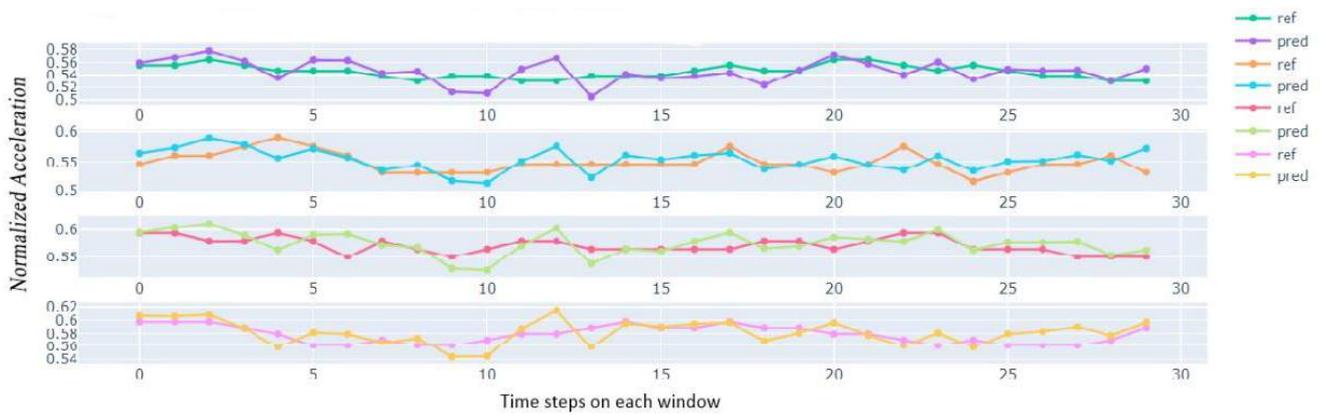

**Fig. 10. Predicted data using a standard neural network.**

### 4- 1- 1- Error Analysis

To have a better understanding of the models' accuracy and optimize the model and obtain better performance in this study, the root mean square error (RMSE) was selected as the loss function of the models in this paper. Furthermore, mean absolute error (MAE) was used as the second metric to show errors. The results are listed in Table 3. For each model, the errors are calculated for training sets and validation sets. As mentioned earlier in this study, the validation sets and training sets are entirely different. Although the MAE depends on the magnitude of outputs, the calculated values can easily be compared due to the same range of output data for different tested models. Increasing the complexity of the model by adding extra convolutional layers reduces both MAE and RMSE. In all cases, the MAE and RMSE for training sets are a bit lower than validation sets, which confirmed a well-fitted model. This can be seen in Fig. 11. It is expected that by increasing the number of faulted sensors with the same number of datasets, the model's errors increase. Although the errors of the model (c) should be higher than the model (a), for the model (c) by overlapping techniques, the number of inputs was increased 15 times larger than the model (a), and this resulted in the lower error.

Fig. 12 shows the convergence curve obtained by training the 3 layers neural network. The loss function, which was the root mean square function (Fig. 12a), and the mean absolute error (Fig. 12b) as other metrics decreased slowly. It is shown that for this case study based on this architecture, after 1000 iterations, validation error did not improve. Therefore, this number of training seems enough.

**Table 3. Errors of the studied cases.**

| Model | No. of faulted sensors | No. of convolutional layers | Datasets | MAE | RMSE |
|---|---|---|---|---|---|
| Model (a) | 1 | 3 | Training | 0.0153 | 0.0210 |
| | | | Validation | 0.0155 | 0.0234 |
| Model (b) | 1 | 2 | Training | 0.0154 | 0.0212 |
| | | | Validation | 0.0161 | 0.0240 |
| Model (c) | 2 | 3 | Training | 0.0106 | 0.0145 |
| | | | Validation | 0.0116 | 0.0162 |
| NN | 1 | 0 | Training | 0.0181 | 0.0235 |
| | | | Validation | 0.0215 | 0.0291 |

MAE: mean absolute error; RMSE: root mean square error

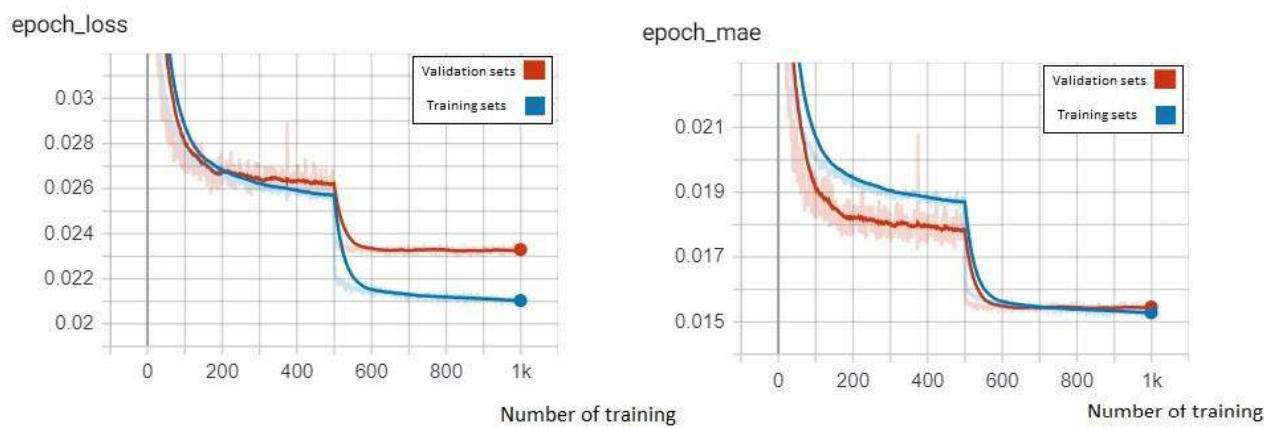

**Fig. 11. Convergence's curve of validation sets and training sets for model (a)**

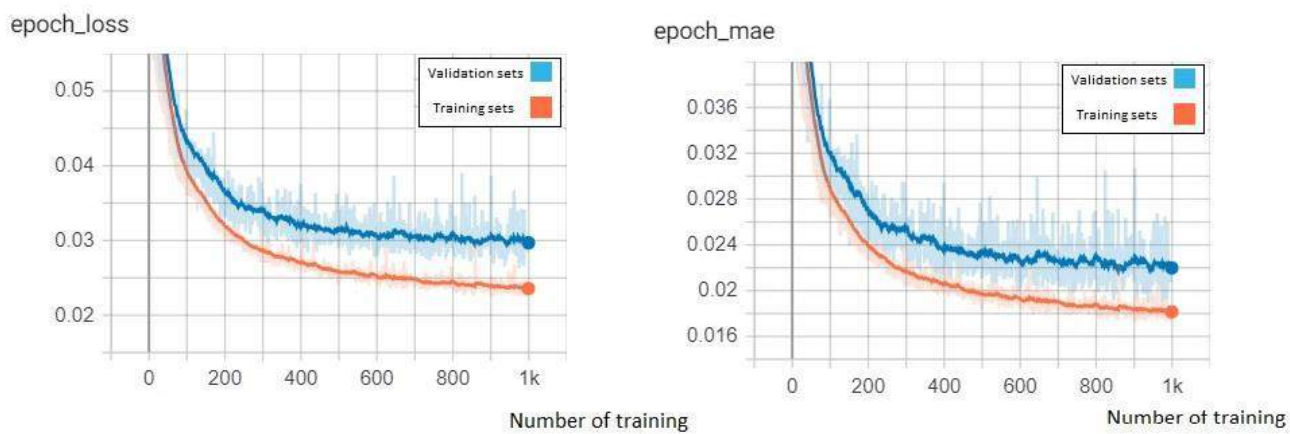

**Fig. 12. Convergence curves for standard Neural Network model.**

**Table 4. Predicted and original natural frequency.**

| Mode | Natural Frequencies (Hz) | | Error (%) |
|------|--------------|-----------|-----------|
|      | Ambient Test | CNN model |           |
| 1    | 7.6210       | 7.7016    | -1.05     |
| 2    | 12.2984      | 12.2446   | 0.43      |
| 3    | 20.1747      | 20.1763   | -0.008    |
| 4    | 24.2876      | 24.2860   | 0.007     |

**Table 5. Modal Assurance Criterion between predicted and original data.**

| Mode | 1     | 2     | 3     | 4      |
|------|-------|-------|-------|--------|
| 1    | 0.904 | 0.237 | 0.016 | 0.002  |
| 2    | 0.178 | 0.979 | 0.001 | 0.0002 |
| 3    | 0.011 | 0.002 | 0.917 | 0.008  |
| 4    | 0.001 | 0.005 | 0.008 | 0.992  |

### 4- 2- Modal Properties

One of the main objectives of structural assessment is to obtain damage-sensitive features from the measured system responses. Among these features, modal properties such as natural frequencies and mode shapes are common. Therefore, to better understand the reliability of the proposed method and parameters to recover lost data, a comparative study was done on the modal features extracted from the predicted CNN model. Hence, a scenario was assumed where a sensor (sensor number 6) misfunctioned. Then trained model (a) was used, and natural frequencies and mode shapes were calculated using the frequency domain decomposition method (FDD) [38], and the results were compared to the original ones. Table 4 listed the natural frequencies

To investigate the correlation between modes from original data and the predicted data (CNN model), Modal Assurance Criterion (MAC), a statistical indicator, was used. Table 5 shows the results.

The MAC is bounded between 0 and 1. The values near one show consistency. Hence, the results in Table 5 showed that the modes shapes from the original signal and predicted signal are consistent. And the CNN model is reliable for predicting data in the SHM application.

### 5- Parametric Study

Another analysis in this study was to investigate the effect of activation functions in hidden layers. The previous results were obtained by using exponential linear unit (eLU) as the activation function in three CNN models and the NN model. Although the nonlinear functions seem to be suitable for CNN, three different nonlinear functions, rectified linear unit (ReLU), eLU, and leaky ReLU were tested for model (a). These activation functions can be calculated, respectively, as follows:

$$f_x = \begin{cases} 0 \ for \ x < 0 \\ x \ for \ x \geq 0 \end{cases} \tag{4}$$

$$f_x = \begin{cases} \alpha(e^x - 1) \ for \ x < 0 \\ x \qquad for \ x \geq 0 \end{cases} \tag{5}$$

$$f_x = \begin{cases} \alpha x \qquad for \ x < 0 \\ x \qquad for \ x \geq 0 \end{cases} \tag{6}$$

Fig. 13 compares the convergence curve of the model (a) for the mentioned activation functions. In this case study, it is concluded that the eLU activation function has a better performance for recovering the data compared to other activation layers. The eLU activation function speeded up the convergence of the model and had a lower error. Increasing the number of epochs reduces the errors of the model. However, after 1000 epochs, the slope of loss function-epochs did not change, and the model did not need further epochs.

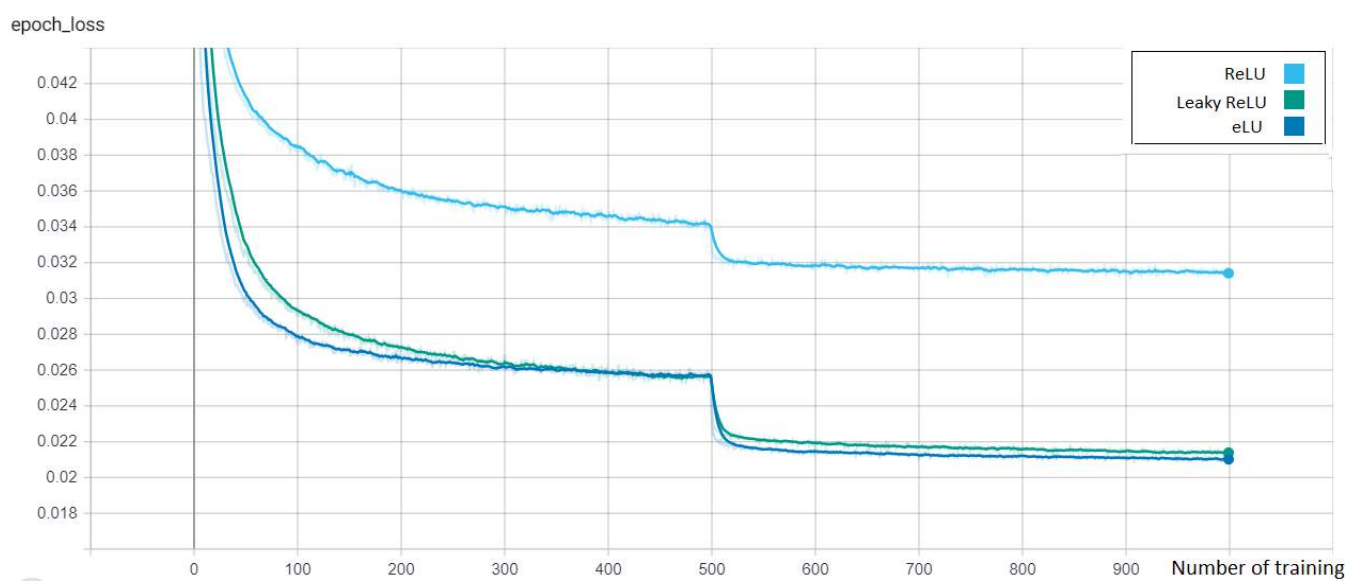

**Fig. 13. Convergence curve of three common activation functions for model (a)**

## 6- Conclusion

In this paper, data sets of a case study on the Alamosa Canyon Bridge were used to predict lost SHM data using a convolutional neural network. Data loss can occur in SHM projects due to data collision in wireless SHM networks or malfunctioned and faulted sensors. The study aimed to validate the accuracy and optimized the performance of the using CNN algorithm as a recovery method for the Alamosa Canyon bridge as an actual structure. The proposed CNN was trained by the measured data of Farrar's experiments [36] on the Alamosa Canyon bridge, which deliberately assumed it had some failed sensors. After training, CNN predicted the missed data of failed accelerometers based on the found correlation of other sensors with high accuracy. Three different CNN architectures are made to evaluate the effects of the number of convolution layers and predicted sensors. The method had a satisfying performance in predicting the missed time histories. As expected, the algorithm had better performance when it had more convolutional layers. Increasing the number of failed sensors can decrease the accuracy of predicted data; however, this error can be reduced by increasing the number of training datasets.

Moreover, the number of datasets plays a pivotal role in improving the efficiency of the algorithm. In SHM of infrastructures like buildings and bridges, the drawback mentioned above is not a serious concern due to the vast amount of raw data; therefore, this method can suitably be employed as a recovery method.

Due to the importance of modal features in damage detection problems and to evaluate the model better, the natural frequencies and mode shapes were obtained using a trained CNN model and were compared to modal features obtained from the original dataset (without any malfunctioned sensors). The results showed that natural frequencies could be predicted with less than 1 %. Error and the predicted mode shapes had an acceptable correlation with the original ones. Also, a neural network was trained to compare the performance of CNN and NN. Based on this study's results, using the mentioned architecture for NN, the algorithm had a fairly well performance without any overfitting. The standard NN error was higher than the proposed CNN; however, CNN's learning speed was approximately two times faster than NN and had a lower computational cost. This is due to the arrangement of data and the advantage of CNN in extracting the features. Therefore, CNN is very efficient in terms of complexity and memory when we have a vast number of data like SHM projects.

**Funding:** The authors received no specific funding for this work.

**Conflict of Interest:** The authors declare that they have no conflict of interest.

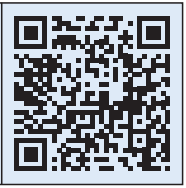